# An Assessment of the UK Government's Clean Energy Strategy for the Year 2030

by
Anthony D. Stephens[1] and David R. Walwyn[2]

**Abstract**

In 2024, the UK Government made two striking announcements on its plans to decarbonise the country's energy system. Firstly, it pledged £21.7 Bn to establish carbon capture and storage hubs on Teesside and Merseyside (Gov.UK, 2024), and secondly it released the Clean Power 2030 Action Plan (UK Government, 2024). The latter outlines a spend of £40 Bn pa for 6 years, a total of £240 Bn, the aim being to decarbonise the electricity system by 2030. The two projects derive from the government's overall ambitions for the UK energy system in 2030, as set out in its 2024 election manifesto (Labour Party, 2024).

This paper questions the validity of both plans, arguing that they do not take adequate account of the consequences of the highly variable nature of wind and solar generations. Using dynamic models of future UK electricity systems which are designed to take account of these variabilities, it is shown that the Clean Power 2030 Action Plan overestimates the ability of wind and solar generations to decarbonise the electricity system as they increase in size relative to the demand of the electricity system. The Action Plan proposes an increase in wind generation from the current level of around 10 GW to around 30 GW, which the dynamic models suggest would result in 26 million tonnes per annum (Mtes pa) of carbon dioxide emissions rather than the 5.2 Mtes pa proposed by the Clean Power 2030 Action Plan. More importantly, the dynamic models show that most of the achievable decarbonisation is the result of increasing wind generation from the current level of around 10 GW to around 20 GW. Increasing wind generation to only 20 GW, rather than to 30 GW as proposed in the Action Plan, should halve the proposed cost, a saving of perhaps £120 Bn, with little disbenefit in terms of reduced decarbonisation.

Dynamic modelling of the wind lull which occurred in January 2017 suggested that, if the wind lull were repeated in 2030, there would be an energy deficit of 4000 GWh. This far exceeds any energy storage which might be envisaged. Also, the wind lull of January 2017 was of only nine days duration, while in previous years there have been wind lulls of much longer duration (Stephens and Walwyn, 2024a). The UK's gas storage capacity of 7.5 winter days looks hopeless inadequate in comparison with the storage capacities deemed necessary by its continental neighbours. As Centrica points out (Centrica Media Centre, 2024), having low gas storage capacity results in gas having to be purchased when it is in short supply and expensive. Centrica's proposal to increase its Rough storage capacity should not only improve energy security but also reduce the cost of both gas and electricity to customers.

Concern is expressed that a consequence of the Climate Change Act of 2008 requiring the UK to meet arbitrary decarbonisation targets is leading government advisors to propose several unproven and therefore highly risky technological solutions. It is suggested that, as was the case prior to the privatisation of the electricity system in 1990, only fully proven technologies should be included in future investment proposals.

---

[1] Correspondence to tonystephensgigg@gmail.com
[2] Graduate School of Technology Management, University of Pretoria, South Africa

1. Introduction

The most complete source of information about the government's energy strategy for 2030 appears in its 2024 election manifesto (Labour Party, 2024), reproduced in Figure 1. The Clean Power Action Plan covers the first four pledges, and the carbon capture and storage pledge appears in pledge 6. Other pledges and the two authorised projects will be discussed later.

To achieve our mission by 2030, a Labour government would:
- Quadruple offshore wind with an ambition of 55 GW by 2030
- Pioneer floating offshore wind, by fast-tracking at least 5 GW of capacity
- More than triple solar power to 50 GW
- More than double our onshore wind capacity to 35 GW
- Get new nuclear projects at Hinkley and Sizewell over the line, extending the lifetime of existing plants, and backing new nuclear including Small Modular Reactors
- Invest in carbon capture and storage, hydrogen, and long-term energy storage to ensure that there is sufficient zero-emission back-up power and storage for extended periods without wind or sun, while maintaining a strategic reserve of backup gas power stations to guarantee security of supply
- Double the government's target on green hydrogen, with 10 GW of production for use particularly in flexible power generation, storage, and industry like green steel

**Figure 1. Extract from the government's 2024 election manifesto showing its energy ambitions for 2030**

2. Clean Power 2030 Action Plan

UK decarbonisation progress is judged by comparing territorial carbon dioxide emissions with those in 1990. Table 2 shows that emissions in 1990 were then about 600 million metric tonnes per annum (Mtes pa), mainly from electricity supply, industry, domestic transport and residential heating. The aim of the Clean Power 2030 Action Plan is to reduce emissions from electricity supply to close to zero in 2030. When the government published its Action Plan in December 2024 it emphasised that it was on the "independent advice" of the National Energy System Operator, NESO. It is therefore to NESO's Clean Power 2030 publication of November 2024 (NESO, 2024) that we must turn for an understanding of the government's Action Plan.

**Table 1. UK carbon dioxide territorial emissions by major sector; 1990 and 2022**

| Sector Emissions (Mtes pa) | Year | |
| --- | --- | --- |
|  | 1990 | 2022 |
| Electricity Supply | 204 | 55 |
| Industry | 156 | 57 |
| Domestic Transport | 129 | 113 |
| Buildings and Product Uses | 108 | 83 |
| Total | 811 | 407 |

Source: Office for National Statistics (2024b)

2.1 NESO's Clean Power 2030 assumptions and predictions

The main assumptions of NESO's Clean Power 2030 publication (NESO, 2024) were:
- an annual electrical demand of 287 TWh (equivalent to wind generation of 29.95 GW)
- nuclear generation of 3.5 GW

- 77.9 GWp[3] of wind capacity (50.6 GWp offshore and 27.3 GWp onshore)
- solar capacity of 47.3 GW.

These assumptions led NESO to advise the government that in 2030 there would be:
- 5.2 Mtes pa of carbon dioxide emissions
- 5.48 GW of excess generation and
- 35 GW of unabated gas turbine generation.

**2.2  Modelling the electricity system anticipated in 2030 using real time historic records**

The Appendix describes two different methods of modelling future UK electricity systems by suitably scaling real time records of wind and solar generations for the year 2017 available on the internet (Gridwatch, 2025).

The model input variables are:
- Hdrm (electrical demand less nuclear generation)
- wm, multiple of wind generation of 6.045 GW in 2017
- sm, multiple of solar generation of 1.16 GW in 2017.

There is considerable uncertainty about the level of nuclear power generation in 2030, and a reasonable assumption is that average Hdrm will be 30 GW in 2030. Assuming offshore and onshore load factors of respectively 43% and 30%, implies wind generation will be 29.95 GW, 4.95 times its value in 2017. The proposed solar generating capacity of 47.4 GW is 3.7 times its 2017. The model input variables for the Clean Power 2030 scenario are therefore:
- Hdrm=30,
- wm=4.95 and
- sm= 3.7.

The Appendix describes the development of a Histogram Model, and Figure 2 shows its application for the Clean Power 2030 assumptions, the blue histogram being the prediction of available wind plus solar generation in 2030 in bands of 5 GW.

Since the electricity system acts towards available wind plus solar generation as a low pass filter, only accepting Hdrm in each generation band,

$$Accommodated\ Generation\ =\ Available\ Generation\ \times \frac{Hdrm}{Generation\ Band}$$

As an example of the application of this algorithm, the available generation in the 62.5 +/- 2.5 GW generation band is 2.5 GW, making the accommodated generation 2.5*30/ 62.5= 1.2 GW. Applying this simple algorithm to all the generation bands generates the red histogram of accommodated wind plus solar generation. The difference between the blue and red histograms is the excess generation which has to be curtailed.

---

[3] Throughout this article, GWp is used to refer to peak capacity, and differs to GW, which refers to the average power output (GWp * efficiency factor).  Typical efficiency factors for offshore wind, onshore wind and solar are 45%, 35% and 25% respectively.

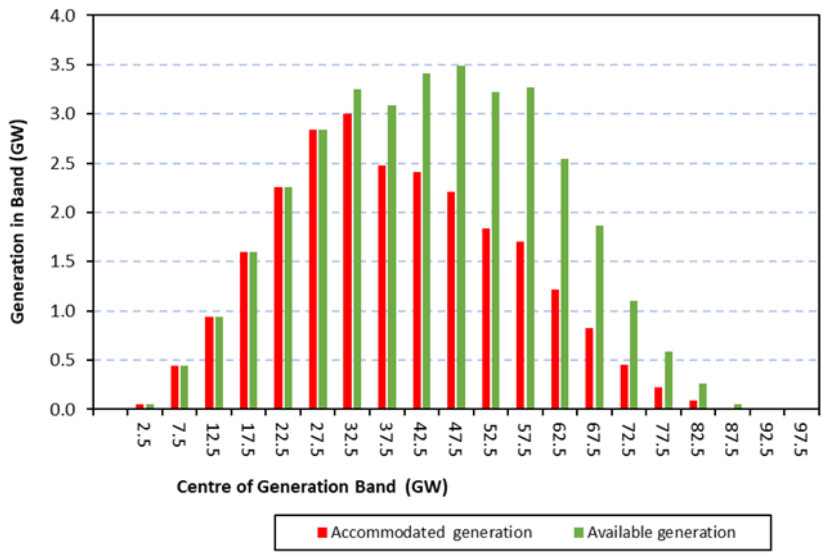

**Figure 2. Prediction of available wind plus solar generation in 2030 in bands of 5 GW (green histogram), and generation accommodated by the electricity system (red histogram), assuming Hdrm = 30, wm = 4.95 and sm = 3.7**

The other modelling approach uses 52 weekly dynamic models and a coordinating model to produce what we have called a Composite Model. This can calculate annual averages, and Table 2 compares the predictions of the Composite and Histogram Models with NESO's predictions of carbon dioxide emissions and excess generation in 2030

**Table 2. Comparison of model predictions for 2030**

| Modelling Approach | Units | NESO (2030 Scenario) | Composite Model | Histogram Model |
|---|---|---|---|---|
| Emissions | Mte pa | 5.2 | 26.0 | 26.2 |
| Excess Generation | GW | 5.48 | 9.48 | 9.66 |

NESO does not explain how it calculated the carbon dioxide emissions, but one possible explanation for the low value of its predictions is that they are based on the four (not dimensioned) "typical" simulations (NESO Figure 17, Figure 3 below), which shows practically no gas generation for much of the year, and therefore little gas emissions.

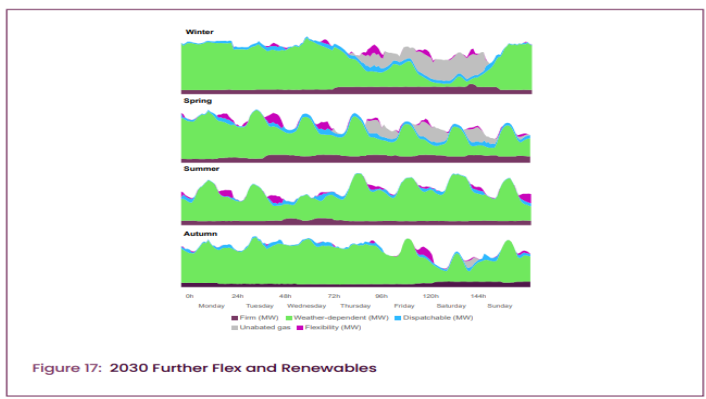

**Figure 3. NESO's illustration of typical weekly gas generation in 2030 (in grey)**

However, the Total Model predicts the 25.96 Mtes of emissions, as may be seen in Figure 4, are spread fairly evenly throughout the year. It is not possible to base predictions on "typical" weeks and expect the results to be meaningful.

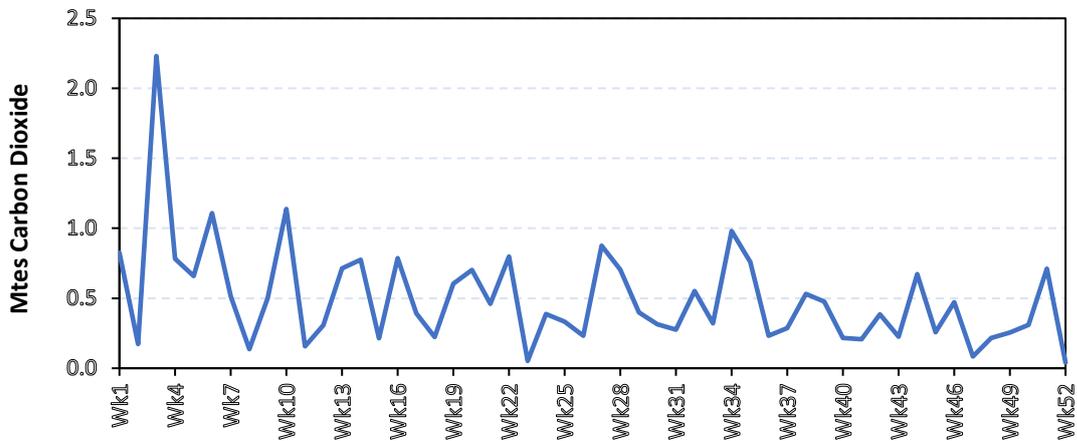

**Figure 4. Prediction of weekly carbon dioxide emissions in 2030 (total 25.92 Mtes pa)**

Since NESO does not explain its calculation methods it is not possible to explain its low predictions in Table 2. However, we suspect that NESOs models take inadequate account of wind and solar variabilities, and therefore of the decreasing efficiency with which they decarbonise the electricity system as they increase in size relative to the demand of the electricity system. The Compound Model is ideally suited to calculating decarbonisation efficiencies, and Figure 3 shows its predictions of carbon dioxide emissions for Hdrm= 30 GW, sm=3.7 (as in the 2030 scenario) and wind multiples from 1 to 10.

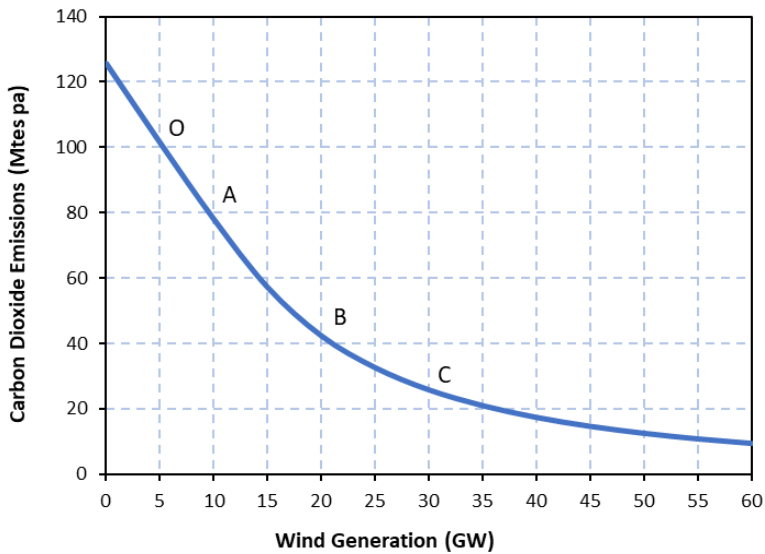

**Figure 5. Compound Model predictions of the relationship between carbon dioxide emissions and wind generation in 2017**

The points highlighted in Figure 5 are:
- O (wm=1, wind generation as in 2017),
- A (wm= 1.65, wind generation 10 GW as in 2024),

- B (wm-3.3, wind generation = 20 GW) and
- C (wm=4.95, wind generation = 30 GW as proposed for 2030).

In addition to using the Compound Model to generate the curve of Figure 5, it was also used to calculate the carbon dioxide emissions for the points O to C highlighted in Figure 3, the results appearing in Table 3. The fifth row in Table 3 show the carbon dioxide emission reductions per GW deployed in moving between adjacent points highlighted in Figure 3. Without any curtailment, a GW of wind would reduce carbon dioxide emissions by 4.87 Mtes pa per GW, and row 6 is the increase in unit cost compared with no curtailment of excess generation.

Table 3 shows little loss in efficiency for wind generation up to around 10 GW, and increasing wind generation from 10 GW to 20 GW reduces carbon dioxide emissions by only 3.56 Mtes pa per GW, resulting in an average unit cost increasing by a factor of 1.37. Increasing wind generation from 20 GW to 30 GW however, results in an average unit cost being increased by a factor of 2.95, which we would suggest is unlikely to be economically unacceptable. Although not included in Table 3, the Compound Model also revealed that 6.58 GW of the curtailed generation of 9.48 GW in Table 2 was the result of increasing wind generation from 20 GW to 30 GW.

**Table 3. Changes to decarbonisation efficiency and energy costs as a function of GWp (wind)**

| Location on Graph | Units | O | A | B | C |
|---|---|---|---|---|---|
| Wind Generation | GW | 6.05 | 10 | 20 | 30 |
| Wind Multiplier (wm) |  | 1.0 | 1.65 | 3.30 | 4.95 |
| Carbon Emissions | Mte pa | 96.35 | 78.08 | 42.47 | 25.97 |
| Incremental Carbon Reduction | Mte pa/GW | 4.87 | 4.63 | 3.56 | 1.65 |
| Unit Cost Multiple |  | 1.00 | 1.05 | 1.37 | 2.95 |

The main conclusion from the above analysis is that most of the benefits in terms of decarbonising the electricity system is gained by increasing wind generation to 20 GW, rather than to 30 GW, as proposed in NESO's Clean Power 2030. If NESO's estimated cost of £240 Bn is correct, this would suggest a potential saving of £120 Bn. Increasing wind generation to only 20 GW should go some way to alleviating Professor Kelly's criticism that there simply aren't the engineering resources available to carry out the project envisaged by NESO (Johnston, 2025).

The conclusion from Table 3 that 20 GW is likely to be the upper economic limit of the wind fleet for an electricity system with a Hdrm of 30 GW is in line with the authors' previous more general finding that wind and solar generation will only be able to decarbonize around 70 % of an electricity system.

## 2.3 Use of Territorial Emissions

It is reasonable to question the validity of the UK monitoring its decarbonisation progress against reductions in "territorial" emissions. Is it because such progress is easy to achieve or perhaps because it allows politicians to virtue signal? As may be seen in Table 1, "territorial" emissions from industry halved between 1990 and 2022. This was not however because industry became more energy efficient but rather because high energy costs caused closure of much of the UK's energy intensive industry. However, in many cases the UK now imports the products it previously manufactured, the emissions now being generated overseas. This was the case with ammonia, which was manufactured on Teesside from 1930 to 2020, and now is being imported into Teesside. The same will be true when INEOS closes its petrochemical manufacture at Grangemouth in 2025 and converts the site to an import terminal. From a climate change perspective, it is only meaningful to monitor emissions if they include both "territorial" emissions and emissions "embedded" in imports

The UK currently monitors decarbonisation progress against territorial emissions only. Particularly since the Department of Environment estimates that emissions "embedded" in imports rose from 34% of the total UK emissions in 1990 to 54% in 2021, to have any validity, strategies should be devised to reduce the UK's total emissions. It is likely that significantly different strategies would result from minimising total emissions, including an increase in manufacturing in the UK (Office for National Statistics, 2024a).

**3.    Teesside Carbon Capture and Storage project**

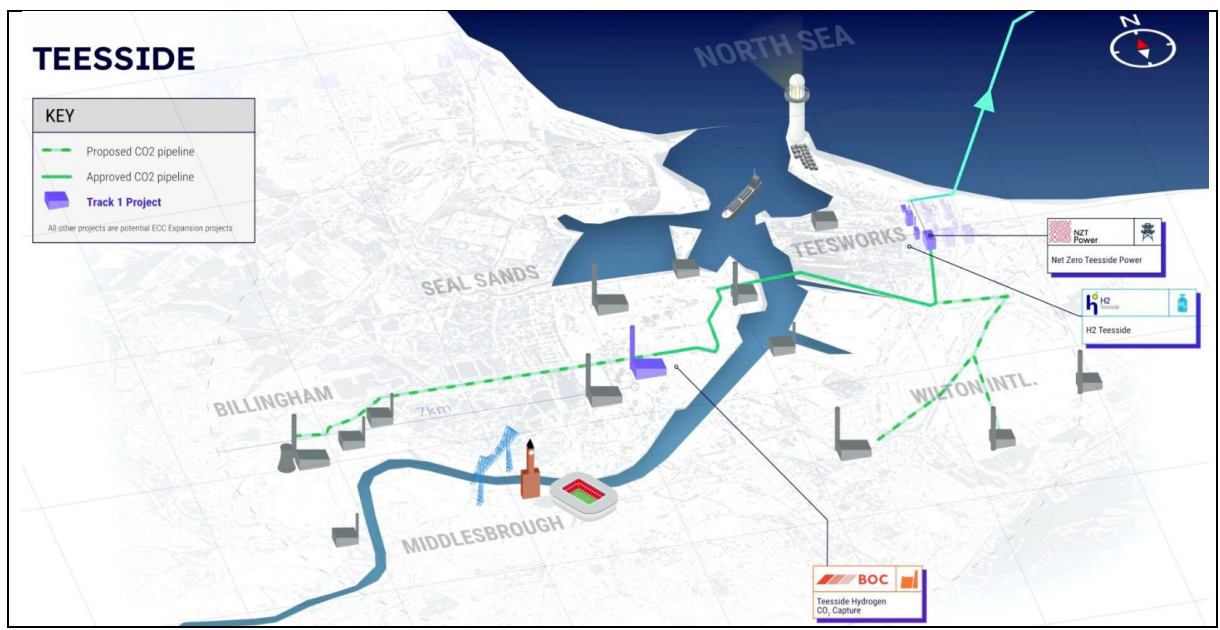

**Figure 6. Teesside Carbon Capture and Storage scheme: carbon dioxide approved pipeline (green), and proposed carbon dioxide pipeline (dotted green)**

Source: Northern Endurance Partnership (2025)

The Teesside carbon capture hub is part of the £21.7 Bn project authorised by the government on 4th October 2025 (Gov.UK, 2024). The initial objective is to capture 4 Mtes pa carbon dioxide from the Track 1 projects highlighted in blue in Figure 6. The carbon dioxide is to be taken 145 km by pipeline to the Endurance carbon dioxide storage repository, whose partners are BP, 45%, Equinor,45%, and Total Energies, 10%. The Net Zero Teesside Power station, which is located to the east of the Tees and close to the sea, is to be a world's first ccgt power station, capturing 2 Mtes of carbon dioxide in generating 742 MW (Net Zero Teesside Power, 2024). The BOC Teesside Hydrogen facility on the other side of the river Tees aims produce 1GW of blue hydrogen by steam reforming natural gas, with carbon capture. Yet to be approved is the 1 GW blue hydrogen facility, H2 Teesside. BPs intention has been to use the hydrogen to provide power to heavy commercial transport in 25 filling stations across the country. It is perhaps surprising that the most advanced project, which will produce 1.2 Mtes pa carbon dioxide, is not to be linked to the carbon dioxide hub. This is the Teesside Renewable Energy plant which will burn 1.1 Mtes pa wood pellets from North Carolina in generating 262 MW; it may be see in Figure 6 adjacent to the inlet on the east side of the river Tees. Its exclusion may be linked to the unresolved controversy about whether burning wood chip is either carbon neutral or sustainable. The 2.6 GW Drax biomass power station is to have its substantial subsidies reduced in future, and its generation restricted to dispatchable generation in winter (BBC News, 2025). It is questionable whether any of the former Billingham and Wilton ICI processes indicated in grey in Figure 6 will actually operate in future to produce carbon dioxide. One of those is the Saudi owned Olefines plant on the Wilton site which has been closed since 2020. Its future operation is likely to depend on its production costs in a highly competitive world marketplace.

In quick succession in February 2025 a number of publications questioned the viability of the Teesside and Merseyside carbon capture and storage proposals for which the government had pledged £21.7 Bn on 4th October 2024 (Gov.UK, 2024)

- on 7[th] February a Commons Public Accounts Committee concluded that there was "*a high degree of uncertainty about whether the risky investment by government will pay off*" (UK Parliament Public Accounts Committee, 2025). It also expressed concern about whether the designs had taken account of recent understanding of liquid natural gas leakages to the environment which *"could undermine the rationale for pursuing certain schemes"*.
- on 14[th] February the Times reported that the *"famously aggressive NY hedge fund Elliott Management"* had amassed a £4Bn position in BP and was pushing for BP to *"ditch green investments"* (Gosden, 2025).
- on 15[th] February an article in the Times suggesting that the *"commitment to subsidise carbon capture and storage"* was making a tempting target for a cash strapped chancellor (O'Connell, 2024) and
- on 16[th] February a Sunday Times article suggested *"We're starting to look like loonies on energy: The rest of the world sees the need for gas. The UK is fuelled by hot air"* (Lawson, 2025).

4. **Security of Supply**

NESO's Clean Power 2030 advice to the government was that in 2030 there will be 35 GW of gas generating capacity, the Labour Party Manifesto making it clear that this capacity is to "guarantee security of supply". However, the House of Lords Science and Technology Committee pointed out in January 2025 that generating capacity does not guarantee security of supply; it is having sufficient gas in storage which keeps the lights on during the worst winter wind lulls (UK Parliament, 2025). The UK experienced a nine day winter wind lull in the 3rd week of January 2017, and several much lengthier wind lulls were experienced in the past (Stephens and Walwyn, 2024a). As explained in the Appendix, the Composite Model includes 52 weekly predictions of the dynamic performance of the future electricity systems based on 2017 records. Figure 7 is the prediction for 2030 should the wind lull of the 3rd week of January 2017 reoccur in 2030.

In January demand and Hdrm are 19% higher than annual average, and the black horizontal line represents the Hdrm of 35.6 GW. The distance between Hdrm and wind plus solar generation must be provided by gas generation, which is a maximum of 33.6 GW on 22nd January. There would not be much safety margin with only 35 GW of gas generation capacity. Although Figure 7 shows that one of the Labour Parties Manifesto pledges was that in 2030 there should be "sufficient zero emission back up power and storage for extended periods without wind or sun", the model shows that the energy deficit during the wind lull of Figure 7 would be 4,000 GWh. In 2022, NG ESO only envisaged 150 GWh of stored energy in 2035, making the manifesto pledge meaningless.

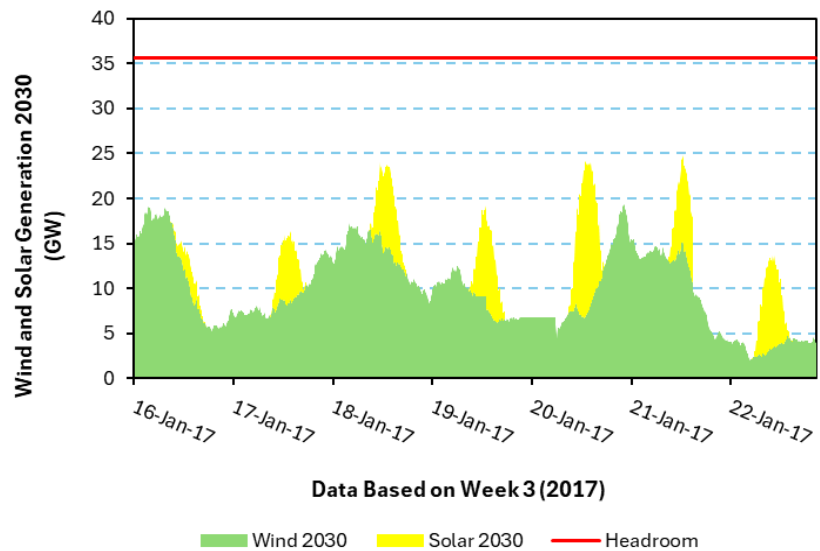

**Figure 7. Prediction for a winter wind lull in 2030 based on wind lull records of January 2017**

The energy company, Centrica, echoed the House of Lords concern about lack of gas capacity, adding that the UK has some of the lowest levels of gas storage in Europe; 12 days average, but 7.5 peak winter days, compared with Germany's 89 days, France's 103 days and the Netherlands 123 days (Centrica Media Centre, 2025). He also pointed out that Centrica was willing to invest £2 billion (of its own money) to increase the capacity of its Rough gas storage facility. This would enable it to buy gas when demand and prices are low, and Centrica estimated that this could have saved consumers £5.2billion over the last two winters (Centrica Media Centre, 2024). The incentive to increase gas storage so that gas purchases can be made when gas prices are low is strongly reinforced by a report in the Economist that "gas sets the price of electricity 98% of the time in Britain compared with a European average of 58%" (The Economist, 2025).

The Institute of Government report on the Energy Market pointed out that, although other generation methods, including renewables and nuclear generation, have lower marginal costs and produce the majority of UK electricity, the price paid to all generators is nearly always the price of gas generation (Bartrum, 2022). This is because energy generation companies bid every half hour for access to the grid, using their marginal costs, but are then paid according to the highest bid, which is normally the gas generation bid. The explanation for this apparently bizarre arrangement is that in any privatised market participants would game the system if paid according to their bids.

The current arrangement of all electricity being priced according to the price of gas has had a devasting impact on heavy industry which was previously a major employer. An elegant solution to this problem has now been suggested (Helm, 2025). Since all renewables and nuclear are contracted to the government, the government could, as in pre-privatisation days, offer electricity to energy intensive industry at marginal cost with a small contribution to fixed costs. This contribution to fixed costs would reduce the cost of electricity for other users.

A major technical risk to the security of future electricity systems is a consequence of the obligation under the Climate Change Act of 2008 for the UK to meet arbitrary decarbonisation targets on pre-determined dates. This has led NESO, and its predecessor NG ESO, to suggest to politicians that they adopt untested technologies, including:

- generation of 5GW then 10 GW of hydrogen by electrolysis, despite heavily subsidised research in the EU and US only achieving only 100 MW
- hydrogen reduced steel, now abandoned by AcelorMittal after building a heavily subsidised prototype because the cost of the resulting steel was too high to be of interest to any of its customers (Collins, 2025)
- steam reforming hydrogen and electricity generation with carbon capture, although there are significant concerns that these processes will generate "fugitive" methane, which will have a significantly more deleterious effect on the environment than carbon dioxide (Alhamdani, Hassim, Ng and Hurme, 2017).

Prior to the privatisation of the electricity supply industry in 1990, promising new technologies were rigorously tested. Only those technologies which were successful when prototyped were then considered for adoption. It is suggested that only technologies which have been rigorously tested, either in the UK or elsewhere, should be included in proposals put to government for investment.

## 5. Conclusion

The paper describes the development of dynamic models of the UK electricity system which take account of the highly variable nature of wind and solar generation. The models provide a means of investigating the decreasing ability of wind and solar fleets to decarbonise the UK electricity system as they increase in size relative electrical demand, and of predicting their economic upper limits. For the UK electricity system suggested by NESO for 2030, the models predict carbon dioxide emissions will be 26 Mte pa, rather than 5.2 Mte pa reported by NESO to the UK government. More importantly, the models suggest that it will only be economic to increase wind generation from the current level of 10 GW to about 20 GW. Should wind generation be increased to 30 GW, only 3.42 GW of the additional 10 GW of available generation would be accommodated by the electricity system, 6.58 GW being curtailed. Restricting wind generation to 20 GW should halve the capital cost, with a potential saving of £120 Bn and contain the unit energy cost within an optimal range.

As explained by the Institute of Government a consequence of the privatisation of the electricity supply industry is that electricity is charged at the price of gas, depriving customers of the benefit of low marginal cost of wind generation (Bartrum, 2022). Helm (2025) has suggested that since all renewables and nuclear are contracted to the government, an elegant solution to this conundrum would be for government to offer electricity to energy intensive industry at marginal cost with a small contribution to fixed costs.

In Section 5.3, Figure 7 illustrated part of a winter wind lull when there was an energy deficit of 4,000 GWh, which is an order of magnitude greater than any stored energy likely to be available in the future. In the period 2013 to 2016, there were no fewer than four occasions with wind lulls of three weeks duration (Stephens and Walwyn, 2024a). It is therefore of some concern that the UK, in stark contrast to other countries, has only 7.5 winter days of gas storage. It would appear that Centrica's proposal to spend £2billionof its own money increasing its Rough gas storage capacity (Centrica Media Centre, 2025) ought to be given higher priority than the £240 billion and £21.7 billion projects to reduce carbon emissions which the Public Accounts Committee deemed to have a "high degree of uncertainty" (UK Parliament Public Accounts Committee, 2025). Not only would this increase the security of the electricity system, but as Centrica explained, it should reduce the price of gas. Since, as explained by the Bartrum (2022), the price of electricity is nearly always determined by the price of gas, increasing gas storage capacity should also reduce the cost of electricity.

Concern is expressed about UK governments being obliged to meet arbitrary decarbonisation targets without proven means of doing so. This has led NESO and its processor NG ESO, to suggest the adoption of several untested technologies. It is recommended that, as prior to the privatisation of the electricity supply industry in 1990, only technologies which have been rigorously tested should be included in investment proposals put to government.

Finally, the paper questions the logic of the UK currently devising strategies to minimise territorial emissions only, particularly since most emissions for which the UK is responsible are now "embedded" in its imports. From a climate change perspective, the objective should be to minimise total emission, i.e. the combination of territorial emissions and emission "embedded" in imports. It is likely that minimising total emissions would result in more manufacturing returning to the UK.

**Appendix. Modelling future UK electricity systems using historic grid records**

The authors have developed two different methods of modelling future UK electricity systems, both based on appropriately scaling historic wind and solar generation records available on the internet (Gridwatch, 2025). Both modelling approaches are based on the finding that annual wind generation histograms for the four years 2013-2016 were similar (Stephens and Walwyn, 2018). This suggested that any of the four years' records might be used to generate future annual wind histograms. Since, solar generation has now become important, and its real time records only became available in 2017, it was decided to predict the behaviour of future electricity systems using 2017 records. As the grid is recorded every five minutes, this required 104,832 wind and solar generation records for 2017 to be downloaded. To ease the problem of handling such a large amount of data, the grid records were downloaded a week at a time, creating 52 weekly models. Each of the wind and solar generation records of 2017 are multiplied by wm and sm, the wind and solar multiples of wind and solar generation in 2017, which were respectively 6.05 GW and 1.16 GW. The scaled records are combined to predict the wind plus solar generation available to the electricity system for each time interval.

A potential modelling complication was that electrical demand varies by typically +/- 10 GW each day and also varies seasonally. A very useful finding was that, if an annual average demand was assumed but modified in the weekly models according to seasonal variations, this had only a second order effect on the annual predictions compared with the prediction of a much more complicated model which used the ever varying electrical demand of grid records downloaded from the internet (Stephens and Walwyn, 2018). Not all electrical demand is available for wind plus solar generation to satisfy, since nuclear generation is given the highest priority access to the grid. We have called electrical demand less nuclear generation the Headroom available to the wind and solar generation, or Hdrm. Dynamic modelling of excess generation shows it to be too intermittent and variable in amplitude to be put to any beneficial use and must be curtailed (Stephens and Walwyn, 2024b). The first public acknowledgement that wind farms were being paid to curtail generation appears to have been in April 2017 (Gosden, 2017). One modelling approach is based on the observation that the electricity system acts as low pass filter towards available wind plus solar generation. This Histogram Model is illustrated in Figure 2, each generation band being tested against the Hdrm. The other approach tests the available wind plus solar generation against Hdrm, for each time interval in the 52 weekly dynamic models. This generates 104, 832 predictions of accommodated wind plus solar generation, which are then averaged to calculate annual averages. We have called the combination of 52 weekly dynamic models and coordinating model a Composite Model.